\documentclass[twocolumn, prl, amsmath, amssymb, amsfonts]{revtex4}

\usepackage[T1]{fontenc}
\usepackage{graphicx}
\usepackage{dcolumn}% Align table columns on decimal point
\usepackage{textcomp}

\usepackage{color}

\renewcommand{\vec}{\boldsymbol}
\newcommand{\db}[2][]{\text{d}^{#1}#2}

\newcommand{\Icl}{\ensuremath{I^\text{cl}}}
\newcommand{\kB}{\ensuremath{k_\text{B}}}
\newcommand{\relint}{\ensuremath{\lambda}}
\newcommand{\sfrac}[1][]{\ensuremath{\gamma_{\text{s}#1}}}
\newcommand{\rhon}{\ensuremath{\rho_\text{n}}}
\newcommand{\rhos}{\ensuremath{\rho_\text{s}}}
\newcommand{\fluct}{\ensuremath{u_r}}
\newcommand{\varfluct}{\ensuremath{\sigma_{u_r}}}

\begin{document}

\title{Mesoscopic Coulomb Supersolid}

\author{A. Filinov}
\email{filinov@theo-physik.uni-kiel.de}
\author{J. B\"oning}
\author{M. Bonitz}
\email{bonitz@physik.uni-kiel.de}
\affiliation{Institut f\"ur Theoretische Physik und Astrophysik, Christian-Albrechts-Universit\"at, Leibnizstr. 15, D-24098 Kiel, Germany}
\author{Yu.E. Lozovik}
\affiliation{Institute of Spectroscopy of the Russian Academy of Sciences, Troitsk, Russia}

\begin{abstract}
When a few tens of charged particles are trapped in a spherical electrostatic potential at low temperature they form concentric shells resembling atoms. These ``artificial atoms'' can be easily controlled by varying the confinement strength. We analyze such systems for the case that the particles are bosons 
and find superfluid behavior which even persists in the solid state. This novel state of matter is a mesoscopic supersolid.
\end{abstract}

%\pacs{}%

\maketitle

%\section{Introduction}

Superfluidity -- the loss of friction due to quantum coherence is well established for bosonic systems in the fluid state. At the same time, experimental observations of superfluidity in solid helium, i.e. of a supersolid \cite{chan04,chan04n}, continue to be controversially discussed, for a recent overview see \cite{balatsky07}. Vanishing of superfluid behavior in slowly annealed helium \cite{reppy07} and restriction of superfluidity to boundaries of crystal grains \cite{sasaki07} raise the question whether at all and if, in what form, superfluidity and solid behavior can coexist. Already the first theoretical predictions \cite{chester67,andreev69,legget70} of a supersolid suggested that this phenomenon may occur only in imperfect crystals. In particular, Andreev and Lifshitz \cite{andreev69} argued that lattice defects (e.g. vacancies or interstitials) are required which may undergo Bose condensation and be responsible for the superfluidity. However, recent simulations showed that in $^{4}$He crystal vacancies tend to phase separate instead of forming a supersolid \cite{boninsegni06}. Some new insight has been obtained by studying small clusters of para-hydrogen \cite{mezzacapo06,ceperley07} which show a substantial amount of superfluidity although the relation of this result to a solid remains unclear. Yet the nature of the supersolid in general and the one in helium in particular remains puzzling \cite{balatsky07}.

It is, therefore, of high interest to analyze superfluidity and supersolid behavior in systems which are more easily controlled than solid helium or para-hydrogen. Examples of model studies include macroscopic systems of hard-core bosons on a triangular lattice \cite{boninsegni05} and dipole interacting bosons in optical lattices \cite{yi07}. Here, we choose a continuous model: a finite number ($N=7\dots 44$) of charged bosons in a spherical parabolic trap in two dimensions (2D) which are known to form crystal-like states consisting of concentric shells \cite{afilinov01}. The advantage of this system is that both, classical and quantum melting are easily externally controlled and crystal symmetry and disorder can be studied in full detail by varying the particle number. We present extensive ab initio path integral Monte Carlo (PIMC) simulations which do not contain any approximation with respect to the interaction, quantum and spin effects. 

Based on our simulations, we report a new state of matter -- a radially ordered mesoscopic crystal which is partially superfluid and will be called ``mesoscopic  Coulomb supersolid''. The crystallization transition is obtained using a novel criterion which is applicable to mesoscopic systems \cite{boening-etal.07}. According to this definition of the ``liquid'' and ``solid''
phases, in mesoscopic solids superfluid fraction can reach up to $\gamma_s=3-4 \%$ for clusters $25 \le N \le 44$ (for $N \le 24$, $\gamma_s$ can be even larger).
Resolving the full superfluid density in radial direction, $\rho_s(r)$, we find that it changes nonmonotonically with the cluster size. In the clusters with the 
configurations $(1,6,\ldots)$ and $(3,9,\ldots)$ of the inner shells we estimate $\gamma_s\approx 3\%$ and the superfluid density, $\rho_s(r)$, is concentrated at the cluster boundaries, i.e the outermost shells.  This is a consequence of the almost hexagonal symmetry of the inner part.
The concentration of $\rho_s(r)$ at the boundary becomes more pronounced for larger clusters.
%, $N=43,44$ where we clearly see that the superfluid behavior vanishes in the inner part close to hexagonal symmetry. The outer shells are strongly disturbed by the trapping potential and contain larger fraction of the full superfluid density. 
Interestingly, for a variety of particle numbers the dominant fraction of the superfluid density is concentrated in the cluster core. These are clusters having 
a structure deviating from hexagonal symmetry (i.e. contain defects), which supports theories predicting the existence of the supersolid phenomenon only for imperfect crystals \cite{chester67,legget70}.

{\bfseries Model and simulation idea.} We consider a two-dimensional system of $N$ identical charged bosons in a harmonic spherical trap with frequency $\omega$ described by the Hamiltonian
\begin{equation}
{\hat H} = - \sum_{i=1}^N \frac{\hbar^2\nabla_i^2}{2m} + \sum_{i=1}^N \ \frac{m\omega^2 r_i^2}{2} + \frac{1}{2}\sum_{i \ne j} \frac{e^2}{|\vec{r}_i-\vec{r}_j|}.
\label{h}
\end{equation}
The thermodynamic equilibrium state of the Coulomb system (\ref{h}) is obtained exactly by using PIMC simulations (see e.g. \cite{afilinov01,cp_book}).
% as reported in \cite{afilinov01}, for details see \cite{cp_book}. 
Each quantum particle is represented by a number of images on $M$  different ``time slices'', and the bosonic spin statistics is fully taken into account by proper symmetrization of the $N$-particle density matrix. To achieve convergent PIMC results for the radially resolved superfluid density we typically used $M=200\dots 500$ and $9 \cdot 10^6$ Monte Carlo steps.

The thermodynamic state is characterized by three parameters: the particle number $N$, temperature and the dimensionless coupling parameter, $\relint=e^2/(l_0\hbar \omega)$ (with the oscillator length $l^2_0=\hbar/m\omega$), which is the ratio of the Coulomb interaction to the trap energy scale. Reducing the trap frequency leads to lower density and increased coupling. In the limit $\relint \rightarrow \infty$ the system approaches a purely classical crystal of point like particles. The simulation results presented below correspond, practically, to the ground state, we use $k_B T/E_0=1/2000$, where $E_0$ is the average Coulomb interaction energy given by $E_0=m\omega^2 r_0^2=e^2/r_0$, where $r_0$ denotes the ground state distance for the case of two particles. 

\begin{figure}[ht]
 \includegraphics[width=0.4\textwidth]{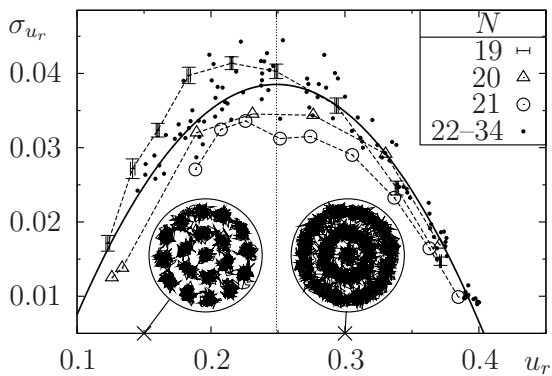}
 \caption{Variance of the mean relative inter-particle distance fluctuations \varfluct{}  versus \fluct{}, Eq.~\eqref{ur}, during the quantum melting transition from a mesoscopic solid (left of peak) to a liquid (right) for $N=19 \dots 34$ particles. The solid black line is a parabolic fit through all data points. The peak position is universal  at $\fluct^\text{crit}\approx0.25$ as indicated by the vertical dotted line, except for $N=19, 29$ where $\fluct^\text{crit}\approx0.22$. Insets show typical configurations for 19 particles at $\fluct=0.15$ and $\fluct=0.3$.}
 \label{fig:1}
\end{figure}

Let us summarize the behavior of the system (\ref{h}). At low density the particles are well localized and arranged in concentric shells, see insets of Fig.~\ref{fig:1}, which are observed in the solid state but also deep into the liquid regime. As for the case of fermions \cite{afilinov01}, shell-structured 2D systems undergo two-stage melting, and two types of order can exist: orientational and radial order (RO). Upon density (or temperature) increase the first transition occurs - orientational melting (OM), i.e. the particles are still localized on shells, but the shells are no more rigidly locked and can rotate relative to each other. 
The second transition - radial melting (RM)  or intershell diffusion happens at much higher densities and temperatures.
In our case the Coulomb crystal melts into a Bose liquid. The phase diagram is very similar to the one of mesoscopic fermions \cite{afilinov01} with two differences. First, the melting point is shifted to lower densities, because Bose statistics enhance particle delocalization in quantum solids and, second, the RO phase, can be partially superfluid.

{\bfseries Detection of crystallization.} While there exists a variety of methods and quantities which allow to determine the melting point (structural transition) in macroscopic systems this is a non-trivial task in a mesoscopic system. Mainly, the difficulties come
from the definition of the term ``phase'' for a system containing several tens of particles and existence of many metastable states,
into which the system can eventually ``jump'' from the ground state. Hence, for mesoscopic systems we cannot define a melting point but a region of crossover where the probability to find the system in one of the metastable states and intermediate transition state become comparable.

Our analysis revealed, that the most reliable and consistent quantity to detect the ``melting'' point is the variance of the block averaged relative inter-particle distance fluctuations $VIDF$ defined as
\begin{equation}
	\sigma_{u_{r}}	= \frac{1}{K}\sum_{s=1}^{K} \sqrt{\langle{u_r^2(s)\rangle}-\langle{u_r(s)}^2\rangle}.
	\label{su}
\end{equation}
where we used the relative interparticle distance fluctuations
\begin{equation}
\fluct(s) = \frac{2}{N(N-1)}\sum_{i > j}^N \sqrt{ \frac{\langle r^2_{ij}\rangle_s}{\langle r_{ij}\rangle_s^2} - 1},
\label{ur}
\end{equation}
which are averaged over sub-interval ``s'' of lenght $M$ of the whole simulation (the full length is $L=M\cdot K$) and
$r_{ij}=|{\vec r}_i-{\vec r}_j|$. 
%However, there is no sharp increase of \fluct{}  during the transition from a mesoscopic solid to a liquid, but only a crossover which extends over a finite range of density (or temperature). Furthermore, in all states fluctuations of \fluct{}  itself are large, as can be seen in Fig.~\ref{fig:1} where we show the variance of \fluct{} , as observed in our simulations, i.e. $\varfluct=[ \langle \fluct^2\rangle - \langle \fluct \rangle^2 ]^{1/2}$. When the system is continuously compressed, \fluct{}  increases from about $0.1$, in the solid state, to about $0.4$, in the liquid. 
Interestingly, the fluctuations \varfluct{} increase non-monotonically in the transition region, having a single maximum. We will use this maximum value of \varfluct{}  as an empirical criterion for separation of the liquid and solid ``phases'', a detailed analysis of this quantity is given in \cite{boening-etal.07}. 

Our results for \varfluct{} for particle numbers $N=19\dots 34$ are shown in Fig. \ref{fig:1} and clearly show this maximum behavior. At the maximum we observe
a critical value $\fluct^\text{crit}\approx0.25$, for most of the clusters, which is surprisingly close to the value $0.249$ following from harmonic lattice theory \cite{baiko} for a macroscopic 3D crystal of charged bosons. Only for $N=19, 29$ the critical value is slightly lower, $\fluct^\text{crit}\approx0.22$. These results will allow us to reliably analyze the behavior of the superfluid fraction in the crossover region and the existence of a mesoscopic supersolid.

\begin{figure}[t]
 \includegraphics[width=0.4\textwidth]{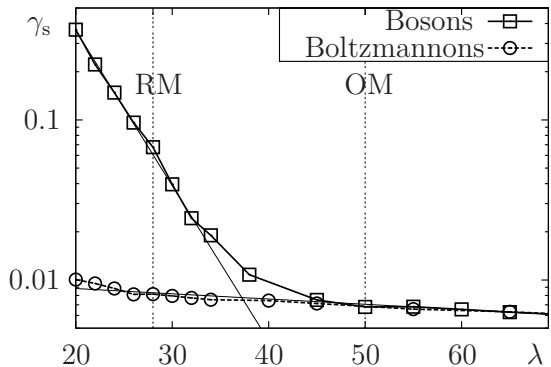}
 \caption{Total superfluid fraction \sfrac{} versus coupling parameter \relint{} for $N=19$. Radial melting (labeled ``RM'') occurs around $\relint\approx 28$ and orientational melting around $\relint \approx 50$. To analyze the systematic error of the computation of \sfrac{} also results for the case of  ``boltzmannons'' are shown.}
 \label{fig:3}
\end{figure}

{\bfseries Superfluid density in the liquid and solid state.} 
Within the two-fluid model the total density is decomposed into a normal and a superfluid component, $\rho=\rhon+\rhos$ with the normalization $\int \rho({\vec r}) \,\db[2]{r} = N$. Usually, one defines the normal component from response of a system to the motion of its boundary. The superfluid fraction \sfrac{}  can be experimentally measured, e.g. Refs.~\cite{chan04,chan04n,reppy07,sasaki07},
or obtained from a numerical experiment~\cite{ceperley99} by measuring the ``missing moment of inertia'' (MMI) related to
the fraction of non-rotating particles
\begin{equation}
\sfrac=\frac{\rhos}{\rho}=\frac{\Icl-I}{\Icl}=\frac{4m^2 \kB T \langle A^2 \rangle}{\hbar^2 \Icl}.
\label{gamma_s}
\end{equation}
Here we consider that the whole system is set into rotation around an axis perpendicular to the 2D plane, in our case, passing through the origin of the trapping potential. Deviation of the quantum, 
$I$, from the classical, $\Icl=\int \rho({\vec r}) mr^2 \,\db[2]{r}$, moment of inertia is related to the average value 
of the square of the total area enclosed by all particle paths in the 2D plane. The instantaneous values $A^2$ for the thermodynamical averaging $\langle \ldots \rangle$ are taken from our numerical simulations of the Bose clusters.

\begin{figure}[t]
 \includegraphics[width=0.4\textwidth]{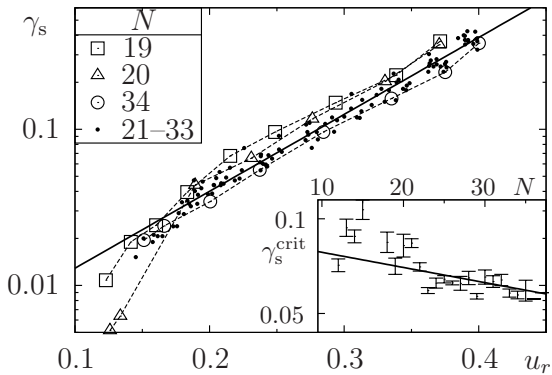}
 \caption{Total superfluid fraction \sfrac{} versus mean relative inter-particle distance fluctuations \fluct{}  for $N=19 \dots 34$ particles (dashed lines are for $N=19, 20, 34$ and scatter points elsewise). The solid black line is the fit, Eq.~\eqref{gamma_fit}, showing the general trend for the liquid regime including partially solid configurations. The inset shows the critical superfluid fraction at the radial freezing point versus cluster size.}
 \label{fig:2}
\end{figure}

The results for \sfrac{} for the cluster $N=19$ are shown in Fig.~\ref{fig:3}. In the liquid state ($\lambda=20$), $\sfrac{} \approx 35\%$, it decreases exponentially with increasing values of the Coulomb coupling parameter \relint. At the melting point (RM) (defined from Fig.~\ref{fig:1})
 $\sfrac\approx 7\%$, decaying rapidly with further increase of \relint. To test the reliability of the results for \sfrac{} we also performed calculations for ``distinguishable quantum particles'' (no exchange), cf. the data labeled ``boltzmannons'' in Fig.~\ref{fig:3}. For this system, in the thermodynamic limit ($N\rightarrow \infty$) the r.h.s of Eq.~(\ref{gamma_s}) asymptotically
goes to zero and $\gamma_s \rightarrow 0$. This is not the case for the simulations with finite $N$, and, formally, we obtain, $\sfrac\approx 0.1 \dots 1 \%$. For ``boltzmannons'', however, this values, rapidly decrease with $N$. The obtained \sfrac{}  can be regarded as a noise level in our computer experiment causing a slight overestimate of the true superfluid fraction.

Systematic deviations from the ``boltzmannon'' curve in Fig.~\ref{fig:3}, confirm that we indeed observe finite superfluidity in the radially ordered (RO) solid phase, $28 \lesssim \relint \lesssim 45$.  For $\relint \gtrsim 45$ the superfluid fraction decays to the noise level of ``boltzmannons''.  Therefore, superfluidity is irrelevant for the  orientational melting boundary (``OM'') which is around $\relint=50$, and the orientational melting is induced by zero-point fluctuations alone, as in the case of fermions~\cite{afilinov01}. The position of this phase boundary is not influenced by quantum statistics. The same trend as shown in Fig.~\ref{fig:3} is observed for different particle numbers. As a result we can conclude that the fully ordered, i.e \emph{radially} and \emph{orientationally}, mesoscopic crystal cannot be superfluid. This is in agreement with the simulations of macroscopic $^4$He solids, e.g.~\cite{boninsegni06}, showing that superfluidity cannot exist in a perfect solid. The peculiarity of the present mesoscopic system is that it has an additional partially (only radially) ordered phase which can support superfluidity.

In Fig.~\ref{fig:2} we summarize the results for the superfluid fraction for all particle numbers in the range from $19$ to $34$. To verify whether there is a universal behavior around the ``RM'' transition we now plot \sfrac{} as a function of the distance fluctuations \fluct{} , using our result [cf. Fig.~\ref{fig:1}] that $\fluct^\text{crit}$ is practically independent of $N$. Indeed, for the present case of quantum melting, a very general trend is observed and, for $\fluct \ge 0.15$, \sfrac{} is well approximated by an exponential 
\begin{equation}
\sfrac(\fluct)=\sfrac^\text{crit} \exp\{\epsilon (\fluct - \fluct^\text{crit}) \},
\label{gamma_fit}
\end{equation}
with the exponent given by $\epsilon=10.68\pm 0.22$ and the critical superfluid fraction at the melting point, $\sfrac^\text{crit} \approx 7.\%$. A more refined analysis yields a weak dependence on the particle number shown in the inset of Fig.~\ref{fig:2} which, for $N\ge 12$, is given by $\sfrac^\text{crit}(N) = c - m N$, where $c=0.090\pm 0.005$ and $m=(8.0\pm 1.7)\cdot 10^{-4}$. This trend indicates that the superfluid fraction in the solid vanishes for cluster sizes around $N=100$. For clusters with commensurable number of particles on neighboring shells, i.e. $N=8, 12, 19, 29$, we find a slightly reduced value of the total superfluid fraction, cf. the points  in the inset of Fig.~\ref{fig:2}. 

%--------------------------------------------------------------------------

{\bfseries Radially resolved superfluidity.}

Since our mesoscopic clusters are strongly inhomogeneous it is tempting to analyze how superfluidity is distributed over the cluster area and what is the influence of the cluster symmetry. 

 There exist different ways to introduce a local superfluid density $\rhos({\vec r})$, e.g. \cite{ceperley07,kwon06}. 
We have found that the normalization \cite{kwon06}
\begin{equation}
\int \rhos({\vec r})mr^2 \,\db[2]{r} = \Icl\sfrac=\Icl-I,
\label{rhos_int}
\end{equation}
is better suited for inhomogeneous  quantum liquids, as it directly characterizes the spatial distribution of the MMI and avoids inconsistencies. The local generalization of \rhos, Eq.~\eqref{gamma_s}, becomes
\begin{equation}
\rhos({\vec r})= \frac{4 m^2 \kB  T \langle A A({\vec r}) \rangle}{\hbar^2 m r^2},
\label{arloc}
\end{equation}
where $A({\vec r})$ is the local contribution to the full projected area $A$ from all path segments contained in the
differential volume $\db[2]{r}$ around point ${\vec r}$.

\begin{figure}[t]
\includegraphics[width=0.4 \textwidth]{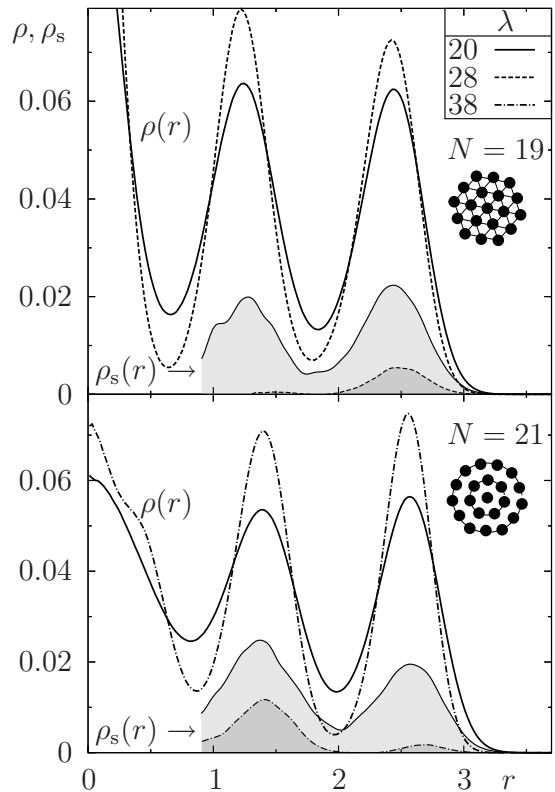}
\caption{Radial distribution of the total density $\rho(r)$ (lines) and the superfluid density $\rhos(r)$ (filled curves) for the magic cluster $N=19$ and the non-magic $N=21$. Both systems are in the liquid phase at $\relint=20$, whereas $\relint=28$ ($N=19$) and $\relint=38$ ($N=21$) correspond to solid-like configurations close to the freezing transition. Insets show ground state configurations of both clusters.}
\label{fig:4}
\end{figure}

Let us first analyze the radial (angle averaged) distribution of the total density $\rho(r)$. In Fig.~\ref{fig:4} we show typical results for $N=19$ and $21$ particles in the liquid ($\lambda=20$) and RO ($\lambda=28,38$) state. In both liquid and
solid states we observe radial modulations of the total density with pronounced deep minima between the shells. The depth of minima increases with $\lambda$, but for Coulomb clusters it changes gradually upon radial freezing in the RO state.

First, we observe that in the liquid state the superfluid density, $\rhos(r)$, and the mass density are equally distributed among the shells, cf. the solid lines for $\relint=20$ in Fig.~\ref{fig:4}. This effect is observed for all cluster sizes $N$ independently of the number of shells and their symmetry.
However, as soon as radial freezing sets in, the spatial distribution of \rhos{} changes dramatically: it is found to be strongly concentrated in different parts of the cluster, in particular on the outer shell ($N=19$) or on the inner shell 
($N=21$), cf. dashed lines in Fig.~\ref{fig:4}. This behavior emerges only in the solid phase and, therefore, must be a consequence 
of the cluster symmetry in the ground state, cf. insets in Fig.~\ref{fig:4}.

Let us analyze the origin of this interesting effect. 
%The nearly equal distribution of the average total density among shells is a well known property of the ground state of Coulomb systems in a harmonic trap. This distribution minimizes the total energy. In the case of charged particles with Bose statistics the system has an additional degree of freedom to reduce the total energy: by concentrating the normal and superfluid density components to different shells. 
The superfluid component which is responsible for the missing moment of inertia is related to quantum coherence of particles which requires overlap of their wave functions. In the liquid phase there is substantial overlap of particles from the same and from different shells. Besides, there are frequent transitions of particles from one shell to another. Consequently, the superfluid component can almost freely ``leak'' from one shell to another which maintains an almost equal peak level of $\rhos(r)$, cf. full lines in Fig.~\ref{fig:4}. On the other hand, in the RO solid phase, particle transitions are strongly suppressed and particle exchanges are rare. 

\begin{figure}[t]
\includegraphics[width=0.4 \textwidth,angle=-90]{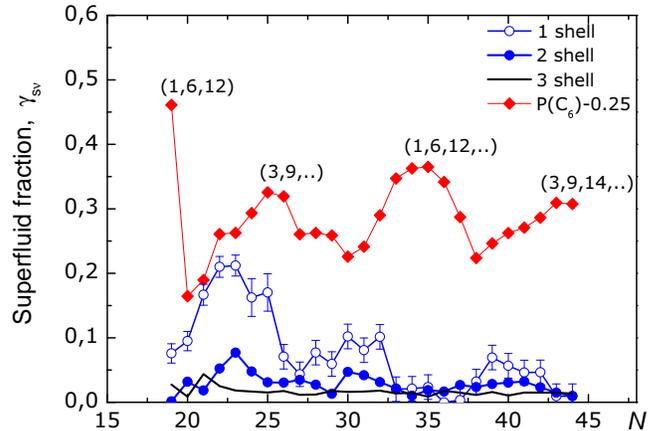}
 \caption{Superfluid fraction $\sfrac[\nu]$, Eq.~\eqref{int_gamma}, of shell $\nu$ vs. particle number in the solid state ($\relint=28)$ compared to the hexagonal symmetry paramter $P_6$ (red rhombs). Starting from $N=31$ a new shell is formed at the center (not shown and counted as $0$-shell).}
 \label{fig:6}
\end{figure}

Consider, for example, the cluster $N=19$. There, in the RO crystal phase, practically the whole superfluid density is concentrated in the outer shell, cf. Fig.~\ref{fig:4}. This is what one would expect from the results for para-hydrogen clusters \cite{ceperley07} or helium grains \cite{sasaki07}: particles at the cluster boundary are less affected by the crystalline order than particles in the bulk and are, thus, the primary candidates for superfluidity. Obviously, this argument holds independently of the particle number:
superfluidity is expected reside in the boundary for clusters which symmetry is close to perfect hexagonal order (which is the symmetry of a macroscopic 2D Coloumb crystal). The clusters with high degree of the hexagonal symmetry are called ``magic'', and $N=19$ is one example.
The shell populations are commensurate (6 particles on the inner and 12 particles on the outer shell), and the whole cluster 
%This is the ``normal'' behavior for any finite piece of crystal lattice as in the case of grain boundaries of $^4$He \cite{sasaki07}. And in fact, 
can be viewed as a small piece of crystal cut out of a macroscopic 2D lattice. As a macroscopic crystal, therefore, for the magic clusters we expect  negligible superfluidity, except for their boundary layer. This, however, describes an idealized situation. For the small clusters considered here, containing only $2-4$ shells, the superfluidity from the boundary layer penetrates into adjacent shells, and they also aquire a finite superfluid density,  cf. Fig.~\ref{fig:6}. However, for the ``magic'' clusters, e.g. $N=19, 29, 34$, the superfluidity on the $2$-nd shells, cf. Fig.~\ref{fig:6}, is significantly suppressed compared to the ``non-magic'' clusters.

In the RO (orientationally disordered) state the definition ``non-magic'' cluster gets a new meaning. Deviations from ``magic'' behavior aris not only from incommensurate shell population but also from a high degree of the orientational disorder of shells.
 In Fig.~\ref{fig:6} we show the $N$-dependence of the $P_6$ -- the probability to find particles with $6$-neighbors on the two inner shells.
The identification of the nearest neighbors was done by performing a Voronoi analysis on every 10-th Monte Carlo step and averaging over ``time slices'' of particle trajectories. In classical magic clusters one finds $P_6$ close to one. 
For quantum clusters, due to quantum-mechanical delocalization of particles, $P_6$ is reduced to values oscillating around $0.5$.
As an example, the cluster $N=30$, has commensurate shell populations ($5,10,15$) but in the RO state it has a value, $P_6 \thickapprox 0.47 $, due to the intershell rotation. To be ``magic'' the cluster should have commensurate shells and be more stable against the shell rotation. For ``magic'' $N=35$, $P_6 \thickapprox 0.60 $.

For ``non-magic'' clusters, as shown in Figs.~\ref{fig:4} and~\ref{fig:6}, superfluidity is concentrated in the cluster core.
To perform a quantitative comparison of the relative amount of superfluid density contained within the shells of ``magic'' and ``non-magic'' clusters and analyze the $N$-dependence, we split the integration in Eq.~\eqref{rhos_int} into non-overlapping shell contributions,
\begin{equation}
\sfrac \Icl = \sum_{\nu} \sfrac[\nu] \Icl_{\nu}, \qquad \Icl_\nu=2\pi \int_{r_{\nu-1}}^{r_{\nu}} \rho(r) mr^2 \,r\db{r}, 
\label{gamma_shells}
\end{equation}
where $r_{\nu-1}$ and $r_{\nu}$ are two adjacent minima of $\rho(r)$ and
\begin{equation}
\sfrac[\nu] = \frac{2\pi}{\Icl_\nu} \int_{r_{\nu-1}}^{r_{\nu}} \rhos(r) m r^2 \,r\db{r}.
\label{int_gamma}
\end{equation}
The results for $N=19 \dots 44$ are shown in Fig.~\ref{fig:6} and confirm the above analysis: while, in the liquid phase, the superfluid fractions on different shells are very close to each other, in the RO solid state, the shells contribute very differently. For clusters with shell occupations $(1,6,12)$ and $(3,9,14)$  we found the highest values of the $P_6$-parameter which proves that they 
have the highest stabity against intershell rotation. As a result the superfluid fraction in the core region (inner shell) is reduced.
This behavior is observed for clusters, $N=19, 26-29, 33-37$ and $43,44$.
In contrast, pronounced concentration of superfluidity in the cluster core is observed for $N=21-23$ and $N=30-32$, $N=39-42$. Again, we observe a close correlation with 
the cluster symmetry given by relatively low values of $P_6$.
Thus, these are the clusters where mesoscopic supersolid behavior is most strongly visible.

Also we can see the general trend that with the increase of $N$, the difference between $\sfrac[\nu]$ on the
$1$-st, $2$-nd and $3$-rd (outermost) shell is reduced. Starting from $N=43,44$, the superfluid fraction at the boundary systematically exceeds that of the inner part. 

The physical mechanism of the present mesoscopic Coulomb supersolid is very different from the one proposed by Andreev and Lifshitz \cite{andreev69} and is not related to the Bose condensation of defects. The source for the supersolid behavior is the partially 
ordered (RO) phase, which is observed in mesoscopic systems in spherical traps. This RO phase, however, vanishes with the increase of the particle number $N$ due to emergence of hexagonal symmetry causing a systematic reduction of the superfluid fraction on the inner shells. 
  
A particular attractive feature of the present system is that supersolid behavior can be externally controlled. By varying the confinement strength the system state can be reversibly changed from partially superfluid ($\lambda < 28 $) to supersolid ($\lambda \gtrsim 28 $). Moreover, with a proper choice of the particle number the superfluid can be directed in a controlled way either to the inner or outer region of the system. Candidates for an experimental observation of mesoscopic supersolid behavior could be bosonic ions in traps or bipolarons in quantum dots.

Financial support by the Deutsche Forschungsgemeinschaft via SFB-TR24 grant A7 and FI 1252 is gratefully acknowledged.

%%%%%%%%%%%%%%%%%%%%%%%%%%%%%%%%%%%%%%%%%%%%%%%%%%%%%%%%%%%%%%%%%

%{\bfseries References and Notes}


\begin{thebibliography}{26}
\bibitem{3d_om} Note that such a state does not exist in three-dimensional crystalline systems \cite{mezzacapo06,ceperley07} due to the much larger inter-shell rotation barrier
\end{thebibliography}
\end{document}